\documentclass[aps,prl, showpacs,twocolumn]{revtex4}
\usepackage{amssymb}
\usepackage{amsmath}
\usepackage{graphicx}
\usepackage{epsfig}

\begin{document}

\title{Perfect transfer of many-particle quantum state via high-dimensional
systems \\
with spectrum-matched symmetry }
\author{Ying Li$^{1}$, Z. Song$^{1,a}$ and C.P. Sun$^{1,2,a,b}$ }
\affiliation{$^{1}$Department of Physics, Nankai University,
Tianjin 300071, China} \affiliation{$^{2}$Institute of Theoretical
Physics, The Chinese Academy of Science, Beijing, 100080, China}

\begin{abstract}
The quantum state transmission (QST) through the medium of
high-dimensional many-particle system is studied with a symmetry
analysis. We discover that, if the spectrum matches the symmetry
of a fermion or boson system in a certain fashion, a perfect
quantum state transfer can be implemented without any operation on
the medium. Based on this observation the well-established results
for the QST via quantum spin chains can be generalized to the
high-dimensional many-particle systems with pre-engineered nearest
neighbor (NN) hopping constants. By investigating a simple but
realistic near half-filled tight-binding fermion system with
uniform NN hopping integral, we show that an arbitrary
many-particle state near the fermi surface can be perfectly
transferred to its translational counterpart.
\end{abstract}

\maketitle

\emph{Introduction.} With minimal spatial and dynamical control
over the interactions between qubits, the transmission of quantum
state through a solid state data bus is an experimental
challenging and a theoretically necessary task for implementing a
scalable quantum computation based on realistic silicon devices.
S. Bose first demonstrated the possibility that use of the quantum
spin system as data bus \cite{Bose1}. In principle, perfect
transfers of quantum state can be implemented by specifically
engineering the coupling constants between the nearest neighbor
(NN) spins in one-dimensional chain \cite{Ekert}. Our recent study
showed that a quantum system possessing a commensurate structure
of energy spectrum matched with the corresponding parity symmetry
can ensure the perfect quantum state transfer also in one
dimensional case \cite{ST}. In this letter we will prove that, to
realize a higher-dimensional QST, a quantum data bus need to
possesses an extendable symmetry matching its spectrum.

The present investigation is motivated by our recent exploration
that an isotropic antiferromagnetic spin ladder system was
proposed as a novel robust data bus \cite{LY}. In such a kind of
gapped system, the non-zero spin states are only virtually
excited, and then the QST has a very high fidelity. It implies
that the perfect quantum-state transfer is also possible in other
high-dimensional many-particle systems as robust quantum data bus.

Actually, many recent researches have been devoted to QST with
quantum chains with permanent couplings or "always-on" inter-spin
couplings
\cite{MYDL04,YUNG2,Bose2,Ekert2,SZ,key-1,TJO04,key-17,key-6,key-18,BBG,PLENIO04}.
Using such kinds of systems as quantum data bus, the quantum
information can be transferred with minimal control only on the
sending and the receiving parties, rather than in the body of data
bus. However most of these mentioned works only focus on the
single-particle and one-dimensional solid state systems. In this
letter a large class of three-dimensional models with modulated NN
interaction is proposed to perform QST. As an application, a
realistic model, that is a simple near half-filled tight-binding
fermion model with uniform hopping integral, is investigated
analytically. It is found that the transmission of the quantum
state near the fermi surface benefits from the quantum correlation
of many-particle system. In other words, for such kinds of models,
the coherent transfer of a many-particle state has higher fidelity
than that of single-particle state at lower temperature.

\emph{Spectrum-symmetry matching condition (SSMC).} To sketch our
central idea, let us first consider a system with the Hamiltonian
$H$ possessing an arbitrary symmetry described by an operator
$R_{q}$, i.e., $[H,R_{q}]=0$. Let $\phi _{n}(q)$ be the common
eigen wave function of $H$ and $R_{q}$ corresponding to the
eigen-values $\varepsilon _{n}$ and $p_{n}$ respectively. It is
easy to find that any state $\psi (q)$ at time $\tau $ can evolve
into its symmetrical counterpart $R_{q}\psi (q)$ if the
eigenvalues $\varepsilon _{n}$ and $p_{n}$ match each other in the
following way
\begin{equation}
\exp (-i\varepsilon _{n}\tau )=p_{n}.  \label{ssmc}
\end{equation}%
We call Eq. (\ref{ssmc}) the spectrum-symmetry matching condition
(SSMC). Actually, an arbitrary state $\psi (q,0)=\sum_{n}C_{n}\phi
_{n}(q)$ at $t=0$ can evolve into
\begin{equation}
\psi (q,\tau )=R_{q}\psi (q,0)
\end{equation}%
at $t=\tau $. Obviously, when the transmission of the quantum
state is concerned, this SSMC always refers to the spatial
symmetry operator, such as the reflection symmetry, translational
symmetry, etc. For the former,
the SSMC appears as so called spectrum-parity matching condition (SPMC) \cite%
{ST}
\begin{equation}
\varepsilon _{n}=N_{n}E_{0},p_{n}=\pm \exp (i\pi N_{n}),
\end{equation}%
for arbitrary positive integer $N_{n}$ and $\tau =\pi /E_{0}$. On
the other hand, when the translational symmetry is concerned, we
have $\exp (-i\varepsilon _{n}\tau )=\exp (-ika)$ where $k$ is
momentum operator and $a $ is translational spacing, i.e., $\exp
(-ika)\psi (i)=\psi (i+a).$ Obviously, it requires $\varepsilon
_{n}=ka/\tau $. An example to demonstrate such kind of scheme is
discussed at the end of this paper. Furthermore the SSMC can be
applied to the high-dimensional many-body system and it should be
helpful to accomplish the task of quantum-state transfer in
practical problems.

\emph{High-dimensional scalable systems for QST.} Consider a
Hamiltonian $H=\sum_{i=1}^{n}H_{i}$ which contains $n$ commutative
sub-Hamiltonians $H_{i}$. The time evolution operator can be
factorized as
$U(t)=\prod_{j=1}^{n}$ $U_{j}(t)$\ by $U_{j}=\exp (-iH_{j}t)$. If $U_{j}$ $%
(\tau )$ is just a certain spatial symmetry operation $R_{j}$ at a
certain instance $\tau $ , such as the reflection or translation,
the time evolution should result in symmetry transformation
$R=\prod_{j=1}^{n}R_{j}$ from the initial state localized around a
spatial point to the final state localized around another point.
The conditions for such high-dimensional scalable systems are two
folds: I. the
spectrum of each $H_{j}$ should satisfy the SSMC; II: the sub-Hamiltonians $%
H_{j}$ should commute with each other. The previous works \cite%
{Ekert,ST,LY,Ekert2,SZ} have proposed many schemes based on
quantum spin system to meet the condition I exactly and
approximately with respect to reflection operation by the spacial
modulation of the NN coupling constants.
\begin{figure}[tbp]
\includegraphics[bb=116 293 506 602, width=6 cm, clip]{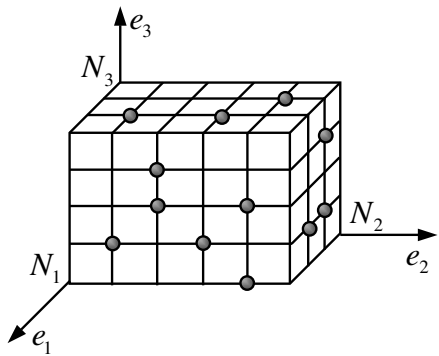}
\caption{\textit{A three-dimensional many-particle system on a
$N_{1}\times N_{2}\times N_{3}$ lattice. It shows that the fermion
or boson system with pre-engineered NN hopping constant can
perform perfect many-particle state transfer.}}
\end{figure}
Now we start our study from a simple system on a three-dimensional
lattice $N_{1}\times N_{2}\times N_{3}$ which is illustrated
schematically in Fig. 1.
\begin{equation}
H=\sum\limits_{i=1}^{3}H_{i}=\sum\limits_{i=1}^{3}\sum\limits_{\mathbf{r},%
\mathbf{\sigma }}J_{\mathbf{r,}i}c_{\mathbf{r},\sigma }^{\dagger }c_{\mathbf{%
r}+\mathbf{e}_{i}\mathbf{,\sigma }}+h.c.,
\end{equation}%
where $c_{\mathbf{r},\sigma }^{\dagger }$ is the fermion or boson
creation operator at
the position $\mathbf{r=}n_{1}\mathbf{e}_{1}+n_{2}\mathbf{e}_{2}+n_{3}%
\mathbf{e}_{3}$ $(n_{i}=1,2,...,N_{i},$ $i=1,2,3)$ with spin $\sigma =\pm 1$%
, $\mathbf{e}_{i}$ is the unit vector for $N_{i}$. The NN hopping constants
are restricted to be nonzero. Obviously, the sub-Hamiltonian $H_{i}$
describes the particle hopping process along the direction $\mathbf{e}_{i}$.
In general cases, the commutation relations between the sub-Hamiltonians are
$[H_{i},H_{j}]\neq 0$.

In the following, we will seek the special formation system with NN
couplings to satisfy $[H_{i},H_{j}]=0$. Notice that for a cubic or square
lattice, this commutation relation can always be rewritten as the form
\begin{equation}
\lbrack H_{i},H_{j}]=\sum_{l}[P_{l},Q_{l}]
\end{equation}%
where
\begin{eqnarray}
P_{l} &=&\sum\limits_{\mathbf{\sigma }}(l_{12}c_{l1,\sigma }^{\dagger }c_{l2%
\mathbf{,\sigma }}+l_{34}c_{l3,\sigma }^{\dagger }c_{l4\mathbf{,\sigma }%
}+h.c.) \\
Q_{l} &=&\sum\limits_{\mathbf{\sigma }}(l_{13}c_{l1,\sigma }^{\dagger }c_{l3%
\mathbf{,\sigma }}+l_{24}c_{l2,\sigma }^{\dagger }c_{l4\mathbf{,\sigma }%
}+h.c.)  \notag
\end{eqnarray}%
denote the sub-Hamiltonian of a $2\times 2$ plaquette labelled by
$l$ (see Fig. 2) and $l_{12},l_{34},l_{13}$ and $l_{24}$ are the
corresponding NN hopping constants. Straightforward calculation
shows that if $l_{12}=l_{34}$ and $l_{13}=l_{24}$ one must have
$[P_{l},Q_{l}]=0$ for any $l$. Therefore, we have reached the
conclusion that $[H_{i},H_{j}]=0$ must hold, if all the
hopping constants $J_{\mathbf{r,}k}$ $(i,j,k=1,2,3)$ are engineered as $%
J_{\mathbf{r}^{\prime }\mathbf{,}k}=J_{\mathbf{r}^{\prime \prime }\mathbf{,}%
k}=...=J_{\mathbf{r}^{\prime \prime \prime }\mathbf{,}k},$ for the sites $%
\mathbf{r}^{\prime },\mathbf{r}^{\prime \prime },...,\mathbf{r}^{\prime
\prime \prime }$ on the the same layer, i.e., $\mathbf{r}^{\prime }\cdot
\mathbf{e}_{k}=\mathbf{r}^{\prime \prime }\cdot \mathbf{e}_{k}=...=\mathbf{r}%
^{\prime \prime \prime }\cdot \mathbf{e}_{k}$. This condition
means that all the hopping constants between two layers ($xy,yz$,
or $zx$) are identical. A simple example is the case that all the
hopping constants along the same direction are identical.
Obviously, the three sub-Hamiltonians commutate with each other.
But we know that the spectrum of such a model does not satisfy the
SSMC.

\begin{figure}[tbp]
\includegraphics[bb=70 350 530 550, width=7 cm, clip]{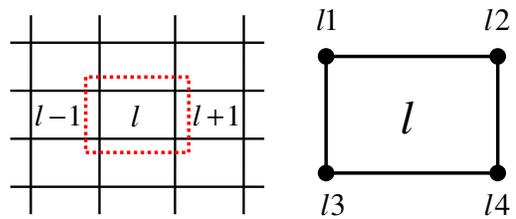}
\caption{\textit{The schematic illustration for the geometry of
the Hamiltonian and its sub block $P_{l},Q_{l}$ on a $2\times 2$
plaquette labelled by $l$.}}
\end{figure}

\emph{Pre-engineered models.} Many people have explored the free
evolution of a single-magnon state via the spin networks to
accomplish QST. It has been found that the homogeneous Heisenberg
and XY spin chain without external field is not a good medium to
transfer a a single-magnon state \cite{ST,SZ}. For a perfect QST
we usually need a pre-engineered spin chain with inhomogeneous
inter-spin couplings \cite{Ekert}. Now we consider a model
Hamiltonian
\begin{equation}
H=\sum\limits_{i=1}^{3}\sum\limits_{\mathbf{r},\mathbf{\sigma }}J_{\mathbf{r,%
}i}^{[m_{i}]}c_{\mathbf{r},\sigma }^{\dagger }c_{\mathbf{r}+\mathbf{e}_{i}%
\mathbf{,\sigma }}+h.c,
\end{equation}%
where the hopping constants are defined as
\begin{eqnarray}
J_{\mathbf{r,}i}^{[m_{i}]} &=&\sqrt{m_{i}[1-(-1)^{n_{i}}]+n_{i}} \\
&&\times \sqrt{m_{i}[1-(-1)^{n_{i}}]+N_{i}-n_{i}}  \notag
\end{eqnarray}%
where $m_{i}$ are arbitrary positive integer numbers. Notice that for fixed $%
m_{i}$ and $N_{i}$, the coupling constants only depend on $n_{i}$,
the label of the layer. Then the pre-engineered
distribution of the hopping constants $J_{\mathbf{r,}%
i}^{[m_{i}]}$ ensures that the sub-Hamiltonians commute with each
other, i.e. $[H_{i},H_{j}]=0$.

To see the dynamic process with such pre-engineered lattice
fermion and boson systems,
we start from the single-particle case by considering one particle state%
\begin{equation}
\left\vert \phi _{\mathbf{k,\sigma }}\right\rangle =\sum_{\mathbf{r}}f_{%
\mathbf{k,r}}\left\vert \mathbf{r,}\sigma \right\rangle \equiv \sum_{\mathbf{%
r}}f_{\mathbf{k,r}}c_{\mathbf{r,}\sigma }^{\dagger }\left\vert
0\right\rangle
\end{equation}%
where $\left\vert 0\right\rangle $ is the vacuum state and $\mathbf{k=}%
\sum_{i=1}^{3}k_{i}\mathbf{e}_{i}$ $(k_{i}=1,2,...,N_{i})$ can be
regarded as the pseudo-momentum, which label the energy levels. Since $%
[H_{i},H_{j}]=0$, the single-prticle wave function can be written
formally as
\begin{equation}
f_{\mathbf{k,r}}=\varphi (k_{1},n_{1})\varphi (k_{2},n_{2})\varphi
(k_{3},n_{3}).
\end{equation}%
From the previous work \cite{ST}, the eigenfunctions of the
sub-Hamiltonian $H_{i}$ with the corresponding eigenvalues
$\varepsilon (k_{i})=-N_{i}+2(k_{i}-m_{i})-1$ for
$k_{i}=1,...,N_{i}/2$; and $\varepsilon
(k_{i})=-N_{i}+2(k_{i}+m_{i})-1$ for $k_{i}=N_{i}/2+1,...,N_{i}$,
can be constructed by the following recurrence equations
\begin{eqnarray}
0 &=&\varepsilon (k_{i})\varphi (k_{i},1)/F(N_{i},m_{i})-\varphi (k_{i},2),
\notag \\
&&....  \notag \\
0 &=&\sqrt{\left( n_{i}+2m_{i}+\Delta \right) \left(
N_{i}-n_{i}+2m_{i}-\Delta \right) }  \notag \\
&&\times \varphi (k_{i},n_{i}+2)+\sqrt{n_{i}+1-\Delta }  \notag \\
&&\times \sqrt{N_{i}-n_{i}-1+\Delta }\varphi (k_{i},n_{i})  \notag \\
&&-\varepsilon (k_{i})\varphi (k_{i},n_{i}+1), \\
&&....  \notag \\
0 &=&\varphi (k_{i},N_{i})-F(N_{i},m_{i})\varphi (k_{i},N_{i}-1)/\varepsilon
(k_{i}).  \notag
\end{eqnarray}%
where $F(N_{i},m_{i})=\sqrt{\left( 1+2m_{i}\right) \left(
N_{i}-1+2m_{i}\right) },$ $\Delta =[1-(-1)^{n_{i}}]/2$.

The above analysis shows that $\left\vert \phi _{\mathbf{k,\sigma }%
}\right\rangle $ is the eigenstate of the total Hamiltonian $H$
with the single-particle spectrum
$E(\mathbf{k})=\sum_{i=1}^{3}\varepsilon (k_{i})$. The spectrums
of one-dimensional systems $\varepsilon (k_{i})$ for $m_{i}=0,1,$
and $2$ are illustrated schematically in the Fig. 3. It shows that
factor $m_{i}$
determines the energy gap between positive and negative energies. For $m_{i}=0$%
, it goes back to the case in the Ref. \cite{Ekert}. This gap
always keep odd times of the minimal level spacing, which plays a
crucial role in the following discussion.

\begin{figure}[tbp]
\includegraphics[bb=60 220 520 655, width=5 cm, clip]{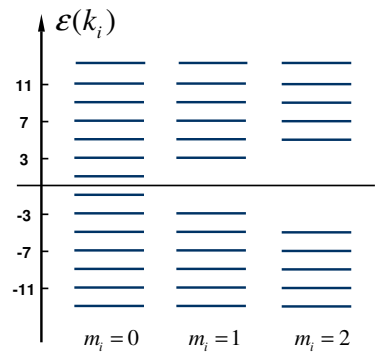}
\caption{\textit{The energy levels for different types of
modulated coupling strength labelled by $m_{i}=0,1,$ and $2$.
Three-dimensional many-body system with such coupling-constant
distribution can perform perfect state transfer.} }
\end{figure}

Defining the diagonal reflection operator $\widehat{R}=\widehat{R}_{1}%
\widehat{R}_{2}\widehat{R}_{3}$, where $\widehat{R}_{i}$ is the
reflection operator with respect to the coordinate about
$\mathbf{e}_{i}$, i.e.,
\begin{eqnarray}
\widehat{R}_{1}\left\vert \mathbf{r,}\sigma \right\rangle &=&\left\vert
\widetilde{n}_{1}\mathbf{e}_{1}+n_{2}\mathbf{e}_{2}+n_{3}\mathbf{e}_{3}%
\mathbf{,}\sigma \right\rangle  \notag \\
\widehat{R}_{2}\left\vert \mathbf{r,}\sigma \right\rangle &=&\left\vert n_{1}%
\mathbf{e}_{1}+\widetilde{n}_{2}\mathbf{e}_{2}+n_{3}\mathbf{e}_{3}\mathbf{,}%
\sigma \right\rangle \\
\widehat{R}_{3}\left\vert \mathbf{r,}\sigma \right\rangle &=&\left\vert n_{1}%
\mathbf{e}_{1}+n_{2}\mathbf{e}_{2}+\widetilde{n}_{3}\mathbf{e}_{3}\mathbf{,}%
\sigma \right\rangle ,  \notag
\end{eqnarray}%
where $\widetilde{n}_{i}=N_{i}+1-n_{i}$ is the
reflection-$\widehat{R}_{i}$ counterpart of the coordinate
$n_{i}$. According to the discussion in Ref. \cite{ST}, we have
$\widehat{R}\left\vert \phi _{\mathbf{k,\sigma }}\right\rangle
=(-1)^{k_{1}+k_{2}+k_{3}}\left\vert \phi _{\mathbf{k,\sigma
}}\right\rangle$. It shows that any minimal change of
$k_{1}+k_{2}+k_{3}$ must induce the change of the parity and
eigenvalue $E(\mathbf{k})$ of the state by the minimal value 2,
simultaneously. Thus it satisfies the SPMC, which guarantees the
perfect state transmission diagonally.

Now we define many-particle state $\left\vert \psi
_{l}\right\rangle =\prod_{\mathbf{k}_{i},\sigma
_{i}}c_{\mathbf{k}_{i},\sigma _{i}}^{\dagger }\left\vert
0\right\rangle $ by the collective operators $c_{\mathbf{k},\sigma
}^{\dagger }
=\frac{1}{\sqrt{\Omega }}\sum_{\mathbf{r}%
}f_{\mathbf{k,r}}c_{\mathbf{r,}\sigma }^{\dagger }\left\vert 0\right\rangle
$
where $\Omega $ is the normalization factor. Obviously, we can have $\widehat{R}%
\left\vert \psi _{l}\right\rangle =(-1)^{p}\left\vert \psi _{l}\right\rangle
$ where $p=\sum_{i}\mathbf{k}_{i}\cdot (\mathbf{e}_{1}+\mathbf{e}_{2}+%
\mathbf{e}_{3})$. Likewise, a straightforward calculation shows
that the state $\left\vert \psi _{l}\right\rangle $ also meets the
SPMC. Then any state $\left\vert \Phi \right\rangle
=\sum_{l}A_{l}\left\vert \psi
_{l}\right\rangle $ at $t=0$ evolves to its reflection counterpart $\widehat{%
R}\left\vert \Phi \right\rangle $ at time $t=\tau =\pi/2$. The
above analysis is available for the fermion and boson systems with
any dimension. But for $XY$ model, this conclusion is true only in
single-particle subspace for $n$-dimensional systems ($n>1$).

\emph{Near perfect transfer in a real many-body system.} Now we
consider how to realize such QST with SSMC in a real fermion
system rather than a pre-engineered system. It is an important
task to seek the practical systems to meet the SSMC approximately.
Most of the previous works focus on single-excitation (-particle)
cases. But the investigation for the QST based on
antiferromagnetic spin ladder \cite{LY} as a robust data bus shows
that many-excitation systems are also good candidate for the task.
On the other hand, the above analysis indicates that the practical
medium could be a many-particle system. To demonstrate this, a
simple example is discussed as following.

\begin{figure}[tbp]
\includegraphics[bb=150 310 470 570, width=6 cm, clip]{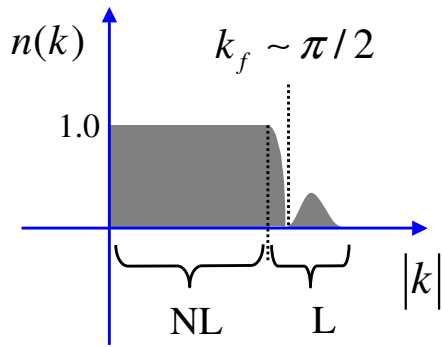}
\caption{\textit{The distribution of particle number in $k$ space
for the states near the fermi surface. The character of such kinds
of states allows us to separate the range of $k$ into linear (L)
and nonlinear (NL) regions.}}
\end{figure}

Consider a simple spinless-fermion model of $\
H=-J\sum_{i}^{N}(c_{i}^{\dagger }c_{i+1}+h.c)$ on an $N$-site ring
with uniform NN hopping constant, In $k$ space, this Hamiltonian
can be diagonalized as $H=-2J\sum_{k}^{N}\cos kc_{k}^{\dagger
}c_{k}$. The single-particle state is not suitable for
implementing QST due to the nonlinear dispersion at lower
spectrum. However, the spectrum in the region around $k=\pm \pi
/2$ has the approximate linear dispersion relation, i.e.,
$\varepsilon (k)\sim -2J|k|+\pi J$. Then, we can
separate the summation of $k$ into two regions: linear $L$ and nonlinear $NL$%
, which is illustrated in Fig. 4 and rewritten Hamiltonian as
\begin{equation}
H\approx -2J\sum_{\left\vert k\right\vert \in
L}|k|n_{k}-2J\sum_{\left\vert k\right\vert \in NL}\cos kn_{k}.
\end{equation}%
It is obvious that a system in the near half-filled case should be
the proper medium for QST at lower temperature. In this case, only
the fermions near the fermi surface are excited and then any
many-particle state can be expanded in terms of the eigenstates of
the form
\begin{equation}
\left\vert \phi _{n}\right\rangle =\prod_{k\in L}c_{k}^{\dagger
}\prod_{k^{\prime }\in NL}c_{k^{\prime }}^{\dagger }\left\vert
0\right\rangle ,
\end{equation}%
with eigen-values $-2J(\sum_{k\in L}|k|+\varepsilon _{0})$, where
$\varepsilon _{0}$ is a constant and $n$ labels the configuration
of the fermions in the linear region. Therefore, near the fermi
surface, arbitrary state will evolve according to the effective
Hamiltonian $H_{eff}=-2J\sum_{\left\vert k\right\vert \in
L}|k|n_{k}+\varpi _{0}$, where $\varpi _{0}$ is a constant which
is independent of the distribution of particle number for $k\in
L$. On the other hand, for an $N$-site ring, the
translational operator $\widehat{T}_{a}$ defined by $\widehat{T}%
_{a}C_{j}^{\dagger }\left\vert 0\right\rangle =C_{a+j}^{\dagger }\left\vert
0\right\rangle $ satisfies
\begin{equation}
\widehat{T}_{a}\left\vert \phi _{n}(0)\right\rangle =e^{i\theta
}\left\vert \phi _{n}(\tau )\right\rangle ,
\end{equation}%
where $\theta$ is independent of the distribution of $k\in L$ and
$\tau =a/2J$. Then as schematically illustrated in Fig. 5, any
state $\left\vert \psi (0)\right\rangle $ localized near the fermi
surface, i.e., which can be expanded by the basis vector
$\left\vert \phi _{n}\right\rangle $ can evolve into its
translational counterpart $\left\vert \psi (\tau )\right\rangle
=\exp (-i\theta)\widehat{T}_{a}\left\vert \psi \right\rangle $ at
instance $\tau =a/2J$.

\begin{figure}[tbp]
\includegraphics[bb=50 396 560 562, width=6 cm, clip]{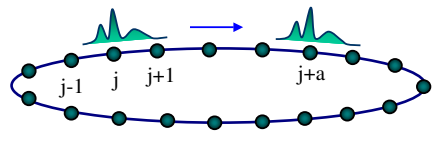}
\caption{\textit{The schematic illustration for the transmission
of the state $\left\vert \psi (0)\right\rangle $ from $j$ to
$j+a$.}}
\end{figure}

\emph{Summary.} In summary, we have extended the SPEC to the
high-dimensional many-body system with other symmetries besides
the parity one. A class of three-dimensional models with modulated
NN interaction is proposed to perform perfect-state transfer in
bulk many-body system. Furthermore, a realistic near half-filled
tight-binding fermion model with uniform hopping integral is
investigated. It is found that, for the many-particle state near
the fermi surface, the eigen values and wave functions satisfy the
SSMC approximately. Then the transfer of a many-particle state has
higher fidelity than that of single-particle state of the same
model, i.e., the many-particle correlation can enhance the
fidelity of the transmission for the corresponding quantum state.

We acknowledge the support of the CNSF (grant No. 90203018, 10474104), the
Knowledge Innovation Program (KIP) of Chinese Academy of Sciences, the
National Fundamental Research Program of China (No. 001GB309310).

\end{document}